\documentclass[]{interact}

\usepackage{epstopdf}
\usepackage[caption=false]{subfig}
\usepackage[numbers,sort&compress]{natbib}
\bibpunct[, ]{[}{]}{,}{n}{,}{,}

\newcommand{\coso}{Cu$_2$OSeO$_3$}

\begin{document}

\articletype{Regular article}

\title{Chiral surface spin textures in Cu$_2$OSeO$_3$ unveiled by soft x-ray scattering in specular reflection geometry}

\author{
\name{V. Ukleev\textsuperscript{a}\thanks{CONTACT V. Ukleev. Email: victor.ukleev@helmholtz-berlin.de, }, C. Luo\textsuperscript{a,b}, R. Abrudan\textsuperscript{a}, A. Aqeel\textsuperscript{b,c}, C. H. Back\textsuperscript{b,c} and F. Radu\textsuperscript{a}\thanks{CONTACT F. Radu. Email: florin.radu@helmholtz-berlin.de}}
\affil{\textsuperscript{a}Helmholtz-Zentrum Berlin f\"ur Materialien und Energie, D-12489 Berlin, Germany; \textsuperscript{b}Physik-Department, Technische Universit\"at M\"unchen, D-85748 Garching, Germany;\\
\textsuperscript{c}Munich Center for Quantum Science and Technology (MCQST), D-80799 M\"unchen, Germany}
}

\maketitle

\begin{abstract}
Resonant elastic soft x-ray magnetic scattering (XRMS) is a powerful tool to explore long-periodic spin textures in single crystals. However, due to the limited momentum transfer range imposed by long wavelengths of photons in the soft x-ray region, Bragg diffraction is restricted to crystals with the large lattice parameters. Alternatively, small angle x-ray scattering has been involved in the soft energy x-ray range which, however, brings in difficulties with the sample preparation that involves focused ion beam milling to thin down the crystal to below a few hundred nm thickness. We show how to circumvent these restrictions by using XRMS in specular reflection from a sub-nanometer smooth crystal surface. The method allows observing diffraction peaks from the helical and conical spin modulations at the surface of a \coso~single crystal and probing their corresponding chirality as contributions to the dichroic scattered intensity. The results suggest a promising way to carry out XRMS studies on plethora of noncentrosymmetric systems hitherto unexplored with soft x-rays due to the absence of the commensurate Bragg peaks in the available momentum transfer range.
\end{abstract}


\section{Introduction}

Recently, noncentrosymmetric magnetic systems attracted attention due to the stabilization of topologically non-trivial spin textures via the antisymmetric Dzyaloshinskii-Moriya interaction (DMI) \cite{bogdanov1994thermodynamically,nagaosa2013topological}. The competition between DMI symmetric exchange, and temperature results in a variety of magnetic field-induced topological spin-swirling textures, such as skyrmions, anti-skyrmions, bi-skyrmions, chiral bobbers, and merons \cite{gobel2021beyond}. The research on magnetic skyrmions has developed into two generic categories, namely, ordered skyrmion lattices in bulk non-centrosymmetric single crystals \cite{tokura2020magnetic}, and  disordered skyrmion structures that occur in various ferromagnetic/heavy-metal multilayers and ferrimagnetic amorphous materials with controlled interfacial DMI that is induced by proximity to layers which exhibit a strong spin-orbit coupling \cite{fert2017magnetic, buttner2018theory, wu2020ferrimagnetic}. The first category offers a versatile platform for studying fundamental collective traits of skyrmions through specific methods such as neutron and x-ray scattering as well as ferromagnetic resonance~\cite{pollath2019}, albeit often in a narrow temperature window that lies below room temperature. The second category emerged as an attractive route for applications in storage media since these skyrmions can be generated and controlled at room temperature in micro-structured devices \cite{luo2021skyrmion}.

$B$20-type cubic chiral magnets like MnSi \cite{muhlbauer2009skyrmion}, FeGe \cite{yu2011near}, FeCoSi \cite{munzer2010skyrmion}, are prototype systems hosting Bloch-type skyrmions, while N\'eel-type ones were observed in polar lacunar spinels GaV$_4$S$_8$~\cite{kezsmarki2015neel}, GaV$_4$Se$_8$ \cite{fujima2017thermodynamically}, and tetragonal VOSe2O$_5$ \cite{kurumaji2017neel}. \coso~is one of the archetypal chiral skyrmion host, which differs drastically from the other $B$20 materials, despite belonging to the same chiral cubic space group $P2_13$ \cite{seki2012observation}. In contrast to the itinerant $B$20s, \coso~is a Mott insulator showing multiferroic properties and proven feasibility to control the skyrmion lattice phase by electric fields \cite{seki2012observation,white2012electric}. Moreover, in contrast to relatively isotropic $B$20-type magnets, \coso~shows a strongly anisotropic phase diagram at low temperatures exhibiting additional tilted conical \cite{qian2018new}, disordered skyrmions \cite{chacon2018observation,bannenberg2019multiple}, and even square and elongated skyrmion phases \cite{takagi2020particle,aqeel2021microwave} when the external magnetic field is applied along one of the cubic crystal axes.
Furthermore, low Gilbert damping in \coso~allows investigations of skyrmion dynamics for high-frequency electronics and spintronics applications in the GHz frequency range \cite{schwarze2015universal,mochizuki2015dynamical}. These unique properties motivate a continuing exploration of \coso.

Resonant elastic soft x-ray magnetic scattering (XRMS) is an unique tool providing element-specific information about long-periodic magnetic structures in single-crystal chiral magnets \cite{mulders2010circularly,yamasaki2015dynamical,zhang2016resonant,ueda2022conical}. The possibility to tune the incident circular or linear polarization of the x-ray beam is particularly exciting for investigation of the topology of magnetic domain walls and winding numbers of the spin-swirling spin textures via circular dichroism  \cite{durr1999chiral,zhang2017direct,zhang2017comms,chauleau2018chirality,legrand2018hybrid}.

Due to the high absorption at soft x-ray energies and limited momentum transfer range, two experimental XRMS geometries are typically utilized: 1) small-angle x-ray scattering from thinned crystals in transmission \cite{yamasaki2015dynamical,okamura2017emergence,ukleev2018coherent,ukleev2019element,burn2019helical,burn2020field,ukleev2020metastable}; and 2) diffraction in the vicinity of a structural (or a magnetic) Bragg peak \cite{langner2014coupled,zhang2016resonant,zhang2017comms,zhang2018reciprocal,zhang2020robust,burn2021periodically}. The latter is only applicable if Bragg reflection is allowed in the momentum transfer range available with the wavelength given at the resonant condition, e.g the $L_3$ or $L_2$ edges of a transition metal. Up to date, the diffraction geometry was mainly utilized to study helical and skyrmion satellites of the anomalous reflection (001) in \coso. Alternatively, for thin films and multilayers the specular \cite{durr1999chiral,zhang2017room,huang2020detection} and multilayer Bragg sheet \cite{chauleau2018chirality,legrand2018hybrid,li2019anatomy,leveille2021chiral} geometries are utilized to detect magnetic satellites of the corresponding reflection. 

So far, detection of the magnetic satellites in a reflection geometry from skyrmionic single crystal hosts has not been demonstrated in the soft x-ray region, mainly due to instruments availability and due to poor surface quality of the single crystals. Here, we show experimentally that for a the sub-nm polished (001) \coso~surface helical and conical magnetic modulations can be observed as satellite reflections around the specular peak (i.e. non-Bragg angle) and that chirality information of the underlying spin textures is encoded as dichroic intensity. This proof-of-principle experiment opens the possibility to explore a plethora of noncentrosymmetric systems hitherto unexplored with XRMS due to the absence of Bragg peaks in the available momentum transfer range at soft x-ray energies.

\section{Materials and Methods}

\begin{figure*}
\begin{center}
\includegraphics[width=1\linewidth]{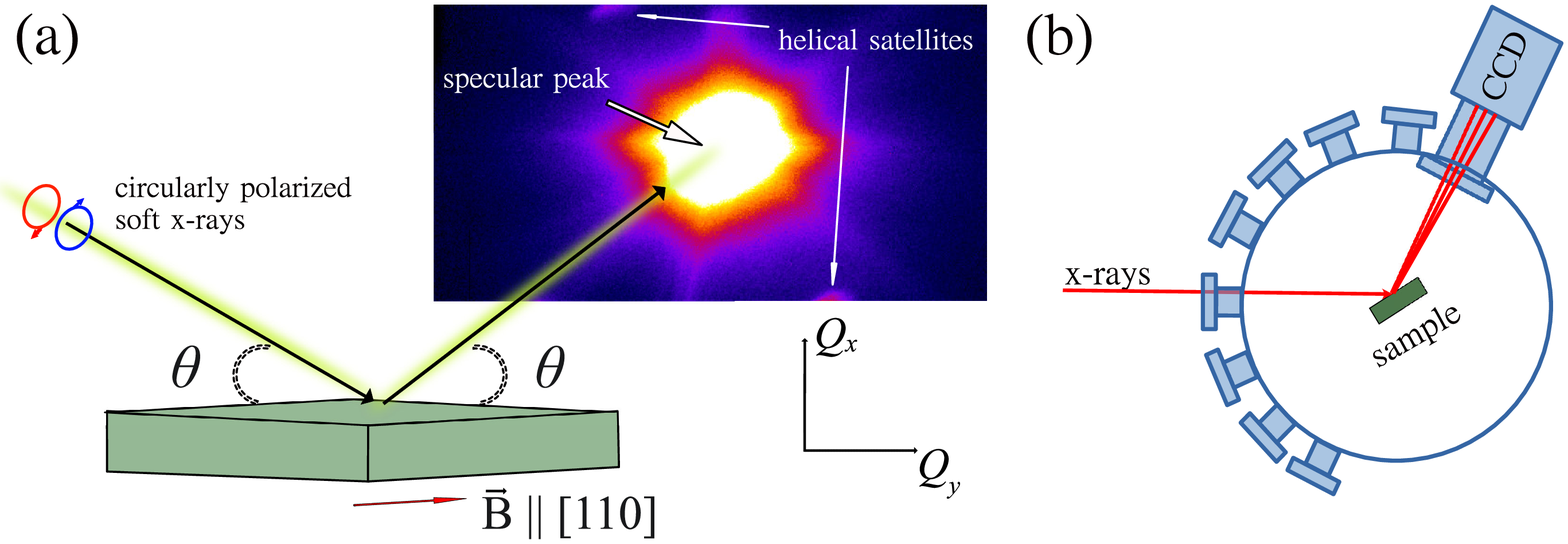}
        \caption{(a) Sketch of the experimental geometry. Circularly polarized soft x-rays impinge onto the samples at an angle $\theta$ and their specular reflection with off-specular magnetic satellites is detected by a position-sensitive 2D detector. The magnetic field $\mathbf{B}$ is applied along the in-plane [110] crystal axis. (b) Top view of the ALICE-II chamber (conceptual draw): the beam enters into the main chamber through a flange that can be selected to meet the required acceptance angle of the CCD detector.}
        \label{fig1}
\end{center}
\end{figure*}

A sketch of the experimental geometry is shown in Fig. \ref{fig1}a. The XRMS experiment was carried out at the dipole beamline PM-3 \cite{kachel2015soft} of the BESSY II synchrotron (Helmholtz-Zentrum Berlin, Germany) using the recently commissioned ALICE-II station dedicated to soft x-ray scattering and coherent diffraction imaging. The main new capability of ALICE-II is that the chamber accommodates a large scattered angle for the CCD detector (up to $2 \theta$=144$^\circ$), as shown in Fig. \ref{fig1}b. This is achieved by choosing the entrance flange that meets the required reflected angle for the CCD. The CCD itself can be mounted at two different distances with respect to the sample, namely at 28\,cm and at 80\,cm, referred to as \textit{low resolution} and \textit{high resolution} options, respectively. Practically, in the \textit{low resolution} setting the CCD chip can accommodate a reciprocal space representing up to $\sim8$\,nm lateral structures at an resonant energy of Cu, and for all available scattering angles.
 
For the current experiments, the chamber was mounted in the low resolution configuration and  two-dimensional (2D) charge coupled device (CCD) detector (Greateyes GmbH, Berlin, Germany) was centered at the scattering angle $2\theta$ of 36$^\circ$. The low resolution and high-$Q$ range option was used. An in-vacuum photodiode and total electron yield (TEY) detectors were used for the sample and chamber alignment with respect to the incident beam direction. The sample was consequently rotated around the $\theta$ axis to access the reciprocal space of interest, around the specular reflection. The base pressure in the chamber was $5\times10^{-8}$\,mbar.

The energy of the circularly polarized soft x-ray beam was tuned to the Cu $L_3$ edge at $E=931.7$\,eV to maximize the magnetic scattering intensity. The incoming beam was collimated by a pair of vertical slits in front of the chamber to reduce the incident beam divergence and hence to improve the angular resolution of the experiment. XRMS patterns were collected at the specular reflection condition ($\theta = 18^\circ$, $2\theta = 36^\circ$) and symmetric off-specular $\pm Q_x$ regions ($\theta = 16^\circ$, $2\theta = 36^\circ$) and ($\theta = 20^\circ$, $2\theta = 36^\circ$) to cover the set of magnetic satellites. The sample was mounted onto a dedicated sample holder and fixed by thermally and electrically conductive silver paint. The magnetic field was applied in the range of $0 - 270$\,mT along the [110] crystal axis by using a rotatable electromagnet. The sample temperature was controlled by a close-cycle cryo-free cryostat (Stinger, ColdEdge Technologies). All measurements were carried out at the base temperature $T=8$\,K. The magnetic field dependencies of XRMS were measured for left and right circularly polarized soft x-rays. The acquisition time for each CCD image was 600\,s. Detector images measured at $B=200$\,mT in the magnetic saturated sample state showing no incommensurate magnetic satellites were used to subtract the background. The aspect ratio of the images in the $Q$-space was corrected to account for the reflection geometry.
 
\begin{figure*}
\begin{center}
\includegraphics[width=1\linewidth]{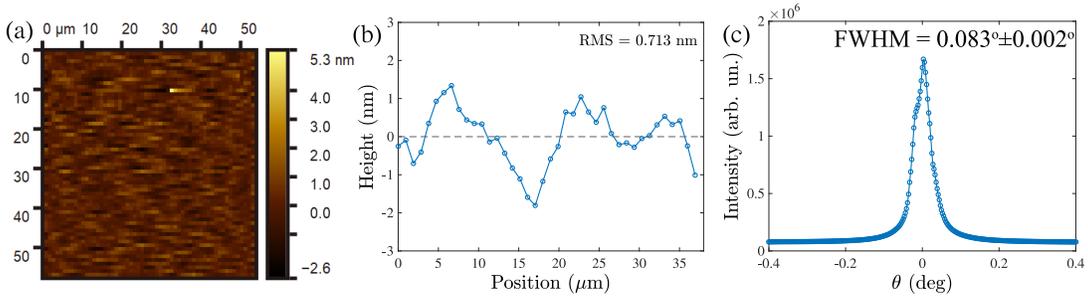}
        \caption{(a) AFM image of the sample surface, demonstrating the exceptionally high quality of the sample's surface (0.7\,nm average roughness) achieved by polishing. (b) Representative line slice of the topography image. (c) Rocking curve measured around the specular peak.}
        \label{fig2}
\end{center}
\end{figure*}

High quality single crystalline \coso~was grown by chemical vapor transport (CVT) \cite{aqeel2022growth}. The crystal was oriented with a Laue diffractometer, cut into a cuboid shape with dimensions $5\times3\times1$\,mm$^3$, and mechanically polished \cite{aqeel2014surface}. The orientation of the polished surface was once again confirmed by x-ray Laue backscattering with sample edges oriented along $\langle110\rangle$~and  $\langle001\rangle$~being along the out of plane crystallographic direction. The surface roughness of the sample was measured by atomic force microscopy (AFM) providing the root mean square value of 7\AA~at the measured area of $50\times50$\,$\mu$m$^2$ (Fig. \ref{fig2}a,b). The surface quality was also confirmed with soft x-rays by means of rocking the sample about the specular reflection (Fig. \ref{fig2}c). The rocking curve shows the full-width at half maxima (FWHM) of the specular reflection peak of $0.083^\circ \pm 0.002^\circ$ confirming the excellent quality of the crystal surface.

\section{Results and Discussions}

\subsection{Field dependence of magnetic satellite peaks in a reflection geometry}

\begin{figure*}
\begin{center}
\includegraphics[width=1\linewidth]{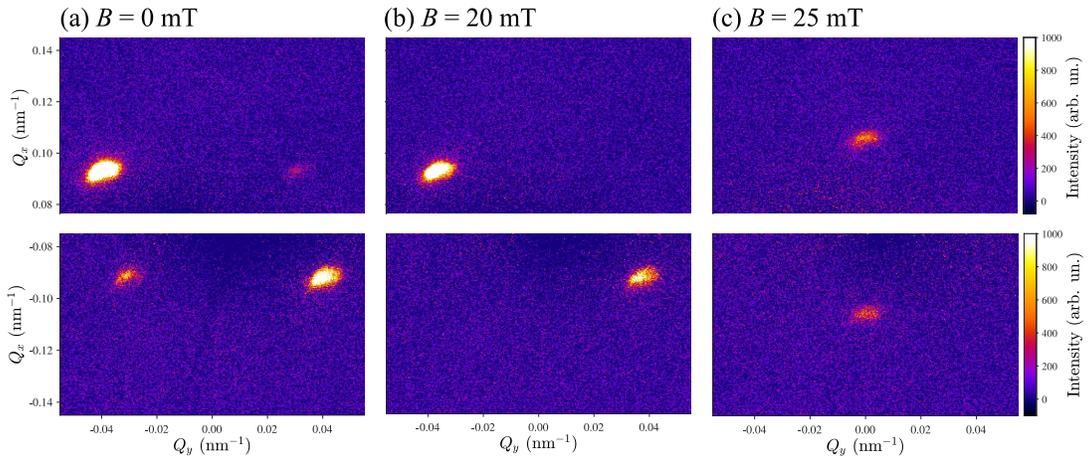}
        \caption{$(Q_x, Q_y)$ maps of the XRMS intensity measured at (a) $B=0$\,mT in the multi-domain helical, (b) $B=20$\,mT in the single-domain helicoidal, and $B=25$\,mT in the conical states.}
        \label{fig3}
\end{center}
\end{figure*}

\begin{figure*}
\begin{center}
\includegraphics[width=1\linewidth]{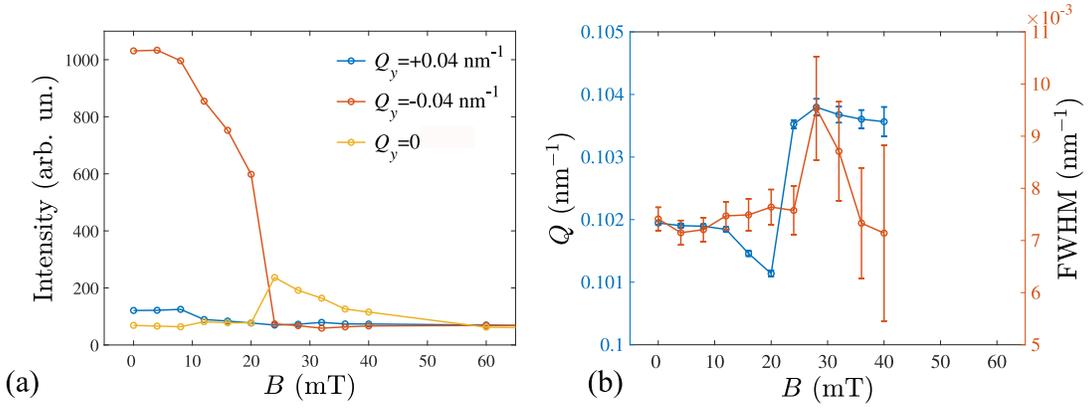}\vspace{3pt}
        \caption{(a) Intensity, (b) position in the $k$-space ($Q$) and width (FWHM) of magnetic Bragg peaks vs. magnetic field dependencies in helical ($0\leq B\leq 20$\,mT) and conical ($20<B\leq40$\,mT) states.}
        \label{fig4}
\end{center}
\end{figure*}

The XRMS patterns measured in the helical ground state at $B=0$\,mT and $T=8$\,K are shown in Fig. \ref{fig3}a featuring four magnetic satellites arising from the proper screw spirals propagating along the in-plane crystallographic directions [100] and [010]. The spiral domain population is unbalanced between the two easy axes due to the magnetic field training applied to the sample during previous scans. Application of a moderate magnetic field of 20\,mT along the middle axis [110] results in the re-orientation of the spiral domains. The re-population takes place due to the field-induced symmetry breaking, which favours the spiral domain with smaller angle between its propagation vector $\mathbf{Q}$ and direction of the magnetic field. Right before the transition to the conical phase at $B=20$\,mT, we observed a fully saturated helical domain state (Fig. \ref{fig3}b). Further increasing the magnetic field results in the transition to the conical state with $\mathbf{Q} || \mathbf{B} || [110]$ (Fig. \ref{fig3}c). Hence, the helical to conical transition is divided into a two step process. This process is reflected in the intensities of the peaks at $Q_y=\pm0.04$\,nm$^{-1}$ showing an interplay in the magnetic field region between 0 and 20\,mT (Fig. \ref{fig4}a). Next, the more intense helical peak re-orients towards the field direction manifesting the first-order helical to conical transition. The magnitude of the spiral wavevector $Q$ decreases as expected from the sine-Gordon law typical for chiral magnetic magnetic solitons (Fig. \ref{fig4}b) \cite{izyumov1984modulated,okamura2017emergence,honda2020topological}. Above the helical-to-conical transition field ($B>20$\,mT) the intensity of the satellite peaks gradually decreases while their position remains unchanged as the magnetization approaches the field-induced ferromagnetic state (Fig. \ref{fig4}). The slight increase of the conical $Q$-vector compared to the helical state (Fig. \ref{fig4}b) is consistent with the previous observation in MnSi \cite{grigoriev2006field}. Interestingly, the widths of the helical and conical peaks are almost constant in the whole field range (Fig. \ref{fig4}b). We note that in the present experiment the width of the satellite peaks is not limited by the instrument resolution or by the specular peak broadening, and reflects the coherent size of helical and conical domains.

\subsection{Chirality sensitivity using circularly polarized light}

\begin{figure*}
\begin{center}
\includegraphics[width=1\linewidth]{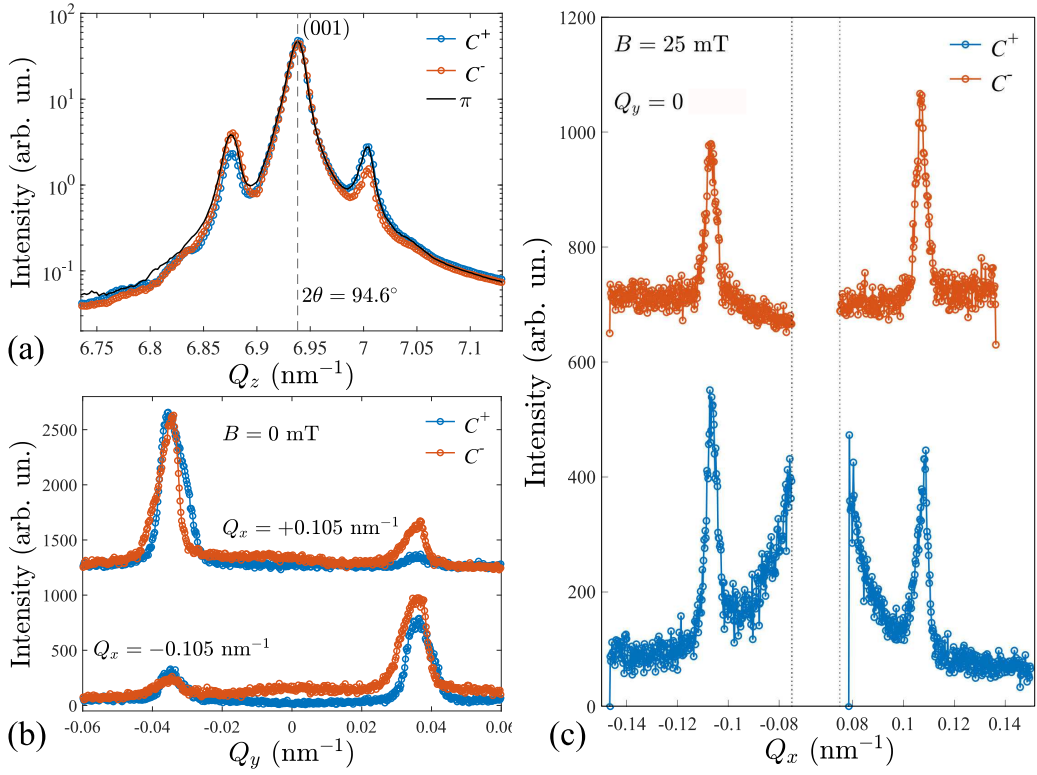}\vspace{3pt}
        \caption{(a) Soft x-ray magnetic scattering (symmetrical scan along $Q_z$, i.e. $\theta$~-~$2 \theta$) in the helical phase, measured for a vanishing external magnetic field and at a temperature of 25\,K. The intensity for the circular positive ($C^+$) and circular negative ($C^-$) x-ray helicity shows a asymmetric behavior on the magnetic satellites of (001) Bragg peak at $2\theta=94.6^\circ$. For linear ($\pi$) x-ray polarization, the intensity of the side peaks are symmetric in intensity. This demonstrates the sensitivity to the chiral nature of the helical phase at the Bragg condition. (b) Line profiles of the specular-XRMS (non-Bragg) intensity along $Q_y$ measured at $Q_x=\pm 0.105$\,nm$^{-1}$ with opposite $C^+$ and $C^-$ polarizations at zero field. (c) Polarization-dependent line profiles of the XRMS intensity along $Q_x$ measured at $Q_y=0$ measured through the conical peaks at $B=25$\,mT.}
        \label{fig5}
\end{center}
\end{figure*}

Chirality of the magnetic structure is an important property of the helical and skyrmion phases. X-ray magnetic scattering with polarized x-rays is an unique tool that can reveal the occurrence of chiral magnetic structure. In Fig. \ref{fig5}a we show XRMS (symmetrical $\theta$~-~$2 \theta$ scans along the $Q_z$ direction) measured around the (001) Bragg peak of the crystal that was set in a helical phase at a temperature equal to 25\,K in our previous experiment \cite{pollath2019}. We probed three basic x-ray polarizations, namely circular positive ($C^+$), circular negative ($C^-$), and linear horizontal polarization ($\pi$). We observed that the intensity of the satellite magnetic peaks do exhibit a strong asymmetry: the intensity of the $C^+$ is lower with respect to the intensity of the $C^-$ for the left side peak and this asymmetry change sign for the right side magnetic peak. The same scan performed with linearly polarized light shows, however, a symmetric intensity behavior. This asymmetry probed by circular light is a direct proof of chirality of the magnetic structure \cite{durr1999chiral} and is fully consistent with the previous observation by Zhang et al. \cite{zhang2016resonant}. Further we show the chirality sensitivity away from the Bragg peak utilizing $C^+$  and $C^-$ circularly polarized light.

Line profiles of the XRMS intensity of helical peaks at zero field along $\mathbf{Q}_y$ measured with two circular $C^+$ and $C^-$ polarizations are shown in Fig. \ref{fig5}b. Intensity ratios between $+Q$ and $-Q$ peaks clearly change as the x-ray polarization is reversed. This is a result of two factors: re-population of helimagnetic domains by the field cycling between the measurements with $C^+$ and $C^-$, and actual circular dichroism of the XRMS intensity due to the topology of the chiral helical magnetic texture in \coso~\cite{zhang2016resonant}. From intensity profiles measured at the same field history (e.g. blue lines in Fig. \ref{fig5}b) it is clear that the intensity ratios between the satellite magnetic peaks located at $Q_y=\pm0.035$\,nm$^{-1}$ differ in the positive and negative intensity slices along $\mathbf{Q}_x$ and hence represent the actual chirality contribution of the XRMS effect, similar to the magnetic scattering around (001) Bragg peak (compare to Fig. \ref{fig5}a).

The intensity difference between $+Q$ and $-Q$ satellite magnetic peaks belonging to the same helical or conical domain is expected due chirality of the magnetic texture. For the present experimental geometry the dichroic contribution to the scattering intensity $I_{CD}=I_{C^-}-I_{C^+}$ reads as \cite{zhang2016resonant}
\begin{equation}
I_{CD} = -Y \sin^2 \xi \sin \theta \cos \psi,
\label{icd}
\end{equation}
where $\xi$ is the conical angle and $\psi$ is the azimuthal angle in the scattering plane ($\psi=0$ corresponds to $\mathbf{Q} || \mathbf{Q}_x$), and $Y$ is a constant determined by the XMCD contrast. In the case of a Bloch-type spin spiral which is characteristic  to chiral magnets, the dichroic intensity follows the cosine law (Eq. \ref{icd}). This equation implies that the reflection geometry ($\theta \neq 0$) is necessary to detect a finite chiral XRMS contribution, while in a transmission experiment the difference between $I_{C^+}$ and $I_{C^-}$ is dominated by other scattering mechanisms such as charge-magnetic interference \cite{eisebitt2003polarization,ukleev2018coherent,ukleev2019element}. Equation \ref{icd} also suggests that for the Bloch-type modulation, the maximum chirality contribution to XRMS is expected at $\xi=0$ and 180$^\circ$ which corresponds to the azimuthal angle of the satellite magnetic peaks in the conical state. Indeed, the intensity switching between $C^+$ and $C^-$ in the conical state ($B=25$\,mT) is found in the line profiles of the XRMS intensity through $Q_y=0$ (Fig. \ref{fig5}d). The dichroic contribution to the intensity is also proportional to the conical angle and hence decreases as the system approaches the field-polarized state. 

\section{Conclusions}

In conclusion, we have demonstrated a proof-of-principle resonant elastic soft x-ray scattering in specular reflection geometry on a polished single crystal of \coso. By measuring the satellite magnetic peaks at a base temperature of 8\,K and as a function of an external field we have observed that the helical magnetic phase evolves from a multi-domain state to a single helical state, right before the eventual onset of the conical phase at $B=20$\,mT. This new transition from a multidomain to a fully saturated single helical domain phase is proposed to originate from the field-induced symmetry breaking which favours energetically the spiral domain  with the propagation vector that is oriented closer to the magnetic field direction. Moreover, we show that chirality information, for both helical and conical phases, is encoded in the asymmetric satellite peak intensities when comparing the scattering cross-sections in a reflection geometry for circular positive with respect to circular negative x-ray beam polarizations. This is of key advantage since the intrinsic chirality of the system cannot be easily reversed through external stimuli.

The method provides a new way to investigate long-periodic spin textures in the near-surface regions of hitherto unexplored non-centrosymmetric magnets with resonant soft x-ray scattering, including depth-sensitive dichroic techniques \cite{zhang2018reciprocal}. Particularly, materials whose lattice parameters do not allow soft x-ray experiments in the Bragg diffraction geometry, such as chiral MnSi \cite{muhlbauer2009skyrmion}, FeGe \cite{yu2011near}, FeCoSi \cite{munzer2010skyrmion}, CoZnMn \cite{tokunaga2015new}, polar GaV$_4$S$_8$ \cite{kezsmarki2015neel}, VOSe$_2$O$_5$ \cite{kurumaji2017neel}, and $D_{2d}$ Mn$_{1.4}$PtSn \cite{nayak2017magnetic}, Cr$_{11}$Ge$_{19}$ \cite{nayak2017magnetic,takagi2018low}, Fe$_{1.9}$Ni$_{0.9}$Pd$_{0.2}$P \cite{karube2021room} are all promising candidates for future structural and dynamics studies using specular geometry reported here. Furthermore, the ALICE-II diffractometer equipped with the 2D detector provides a great opportunity for the real-space imaging in reflection geometry by exploiting coherence of modern synchrotron undulator beamlines.

\section*{Acknowledgement(s)}
We would like to thank Torsten Kachel (BESSY II) for his help with the beamline operation.

\section*{Disclosure statement}

Authors declare no conflict of interest.

\section*{Funding}

We acknowledge financial support for the VEKMAG project and for the PM2-VEKMAG beamline by the German Federal Ministry for Education and Research (BMBF 05K2010, 05K2013, 05K2016, 05K2019) and by HZB. F.R. acknowledge funding by the German Research Foundation via Project No. SPP2137/RA 3570. This work has also been funded by the Deutsche Forschungsgemeinschaft (DFG, German Research Foundation) under SPP2137 Skyrmionics and the excellence cluster MCQST under Germany’s Excellence Strategy EXC-2111 (Project No. 390814868).

\bibliographystyle{tfq}

\end{document}